\documentclass[aps,pre,twocolumn,preprintnumbers,amsmath,amssymb,amsfonts,superscriptaddress, floatfix]{revtex4-1}
\usepackage{color}
\usepackage{xcolor}
\usepackage{mathrsfs}
\usepackage{float}
\usepackage{esint}
\usepackage{bm}
\usepackage{graphicx}
\usepackage{amsmath}
\usepackage[normalem]{ulem}

\renewcommand\phi{\varphi}

\newcommand{\Q}{\mathbf{Q}}
\newcommand{\bu}{\mathbf{u}}
\newcommand{\n}{\mathbf{n}}

\begin{document}

\title{Active nematic defects in compressible and incompressible flows}
\author{Supavit Pokawanvit}
\affiliation{Department of Applied Physics, Stanford University, Stanford, CA 94305}
\author{Zhitao Chen}
\affiliation{Department of Physics, University of California Santa Barbara, Santa Barbara, CA 93106}
\author{Zhihong You}
\affiliation{Fujian Provincial Key Laboratory for Soft Functional Materials Research, Research Institute for Biomimetics and Soft Matter, Department of Physics, Xiamen University, Xiamen, Fujian 361005, China}
\author{Luiza Angheluta}
\affiliation{Department of Physics, University of Oslo, P. O. Box 1048, 0316 Oslo, Norway}

\author{M.~Cristina Marchetti}
\affiliation{Department of Physics, University of California Santa Barbara, Santa Barbara, CA 93106}

\author{Mark J. Bowick}
\email{bowick@kitp.ucsb.edu}
\affiliation{Kavli Institute for Theoretical Physics, University of California Santa Barbara, Santa Barbara, CA 93106}

\begin{abstract}
We study the dynamics of active nematic films on a substrate driven by active flows with or without the incompressible constraint.Through simulations and theoretical analysis, we show that arch patterns are stable in the compressible case, whilst they become unstable under the incompressibility constraint. For compressible flows at high enough activity, stable arches organize themselves into a smectic-like pattern, which induce an associated  global polar ordering of $+1/2$ nematic defects. By contrast, divergence-free flows give rise to a local nematic order of the $+1/2$ defects, consisting of anti-aligned pairs of neighboring defects, as established in previous studies.
\end{abstract}

\maketitle

\section{Introduction}
\label{sec:intro}
Active nematic liquid crystals are fluids of elongated entities that exert forces on their environment, driving self-sustained flows~\cite{doostmohammadi2018active}. Nematic order is found ubiquitously in active systems, from suspensions of cytoskeletal filaments and associated motor proteins~\cite{sanchez2012spontaneous,zhang2018interplay} to bacteria~\cite{thutupalli2015directional,doostmohammadi2016defect} and epithelial cells~\cite{duclos2017topological,blanch2018turbulent}. At high enough activity, all these systems exhibit spontaneous spatio-temporal chaotic flows  sustained by the balance between energy input at the microscale and energy dissipation via internal viscous stresses or friction with the environment. This rich dynamics is controlled by the energy exchange between flow and the elastic liquid crystalline degrees of freedom that control the nematic texture. In the chaotic dynamical state, turbulent-like vortical flows are accompanied by the generation of unbound half-integer topological defects in the nematic texture~\cite{alert2022active}. Defects in active nematics have been at the center of attention for some time. They play a role in driving and controlling flows~\cite{giomi2013defect,pismen2013dynamics,giomi2014defect,keber2014topology}, have been suggested to have possible biological functions~\cite{kawaguchi2017topological,saw2017topological,maroudas2021topological,copenhagen2021topological}, and can themselves organize in striking ordered structures~\cite{doostmohammadi2016stabilization}. Key in controlling defect dynamics and organization are the active flows generated by the distortions in the nematic texture.

In this paper we examine the role of flow incompressibility in controlling the emergent states of active nematics in two dimensions. We do so by comparing two minimal models of extensile active fluids, both at constant density, and where dissipation is controlled entirely by frictional coupling to a substrate. In the first model, we eliminate pressure by imposing incompressibility on the flow. In the second, we do not require incompressibility, but assume the density is kept constant, for instance via birth and death processes, resulting in vanishing pressure gradients. The resulting minimal model, referred to as compressible, exhibits emergent patterns qualitatively similar to those obtained in ``dry'' systems with density fluctuations~\cite{chate2020dry}. We show that even without viscous stresses, incompressibility induces long-range constraints in the flow that strongly alter the collective behavior of nematic defects.
In the absence of incompressibility and with increasing activity, {flow-aligning ($\lambda>1$) extensile nematics transition} through the well-established bend instability~\cite{simha2002hydrodynamic} to a state of aligned arches of the director field or kink walls of the order parameter field~\cite{srivastava2016negative, putzig2016instabilities, patelli2019understanding}. 
The arches form a smectic-like structure and guide the defect dynamics, driving polar order of $+1/2$ defects which travel along the arch walls, ``unzipping" the nematic director field. These structures have been predicted before~\cite{shankar2019hydrodynamics} and observed in simulations~\cite{putzig2016instabilities,srivastava2016negative,patelli2019understanding}. 
Evidence of aligned arches has also been found in strongly damped cell layers~\cite{duclos2017topological}. We demonstrate the linear stability of the arch states for compressible flow through an analytical calculation.
When the flow is incompressible, in contrast, arches are unstable and the system transitions directly from the bend state to active turbulence. 
In this case, $+1/2$ defects exhibit only local nematic order, corresponding to  anti-aligned neighboring defect pairs. 

The remainder of the paper is organized as follows. We  introduce the two models in Sec.~\ref{sec:hydro}, then present the results of numerical simulations in Sec.~\ref{sec:numerics}, including phase diagrams in terms of activity.  In Sec.~\ref{sec:KW}, we perform analytically the linear stability of the arch state and demonstrate the key role of incompressiblity in destabilizing arches. Statistical properties of defect ordering are discussed in Sec.~\ref{sec:order}. We conclude in Sec.~\ref{sec:conclusion} with a brief discussion.

\section{Hydrodynamic model}
\label{sec:hydro}
We consider a two-dimensional active nematic film on a substrate described by a flow field $\mathbf{u}$ and an orientational order parameter tensor $Q_{ij}=S(n_i n_j -\frac{1}{2}\delta_{ij})$.
Here, $S$ is a scalar that quantifies the magnitude of orientational order and the director $\n = [\cos\theta, \sin\theta]$  is a headless unit vector that identifies the direction of spontaneously broken rotational symmetry. The evolution of the  order parameter is given by~\cite{beris1994thermodynamics,marenduzzo2007hydrodynamics,doostmohammadi2018active} 
\begin{equation}
\left(\partial_t+ \mathbf{u}\cdot\bm \nabla\right)\Q+\bm\omega\cdot\Q-\Q\cdot\bm\omega=\lambda\mathbf{D}+\frac{1}{\gamma}\mathbf{H}\;,
\label{eq:Qeq}
\end{equation}
where $\mathbf{D}=\frac12(\bm\nabla\mathbf{u}+\bm\nabla\mathbf{u}^T-\mathbf{1}\bm\nabla\cdot\mathbf{u})$ and $\bm\omega=\frac12(\bm\nabla\mathbf{u}-\bm\nabla\mathbf{u}^T)$ are the traceless symmetric  and antisymmetric parts of the strain rate tensor, respectively. 
The flow alignment parameter $\lambda$ is controlled by molecular shape and degree of orientational order. Here we consider a fluid of uniaxial elongated nematogens where {$\lambda>1$}. The molecular field $\mathbf{H}$ controls the relaxation dynamics, with $\gamma$ a rotational viscosity. It is obtained from a Landau-de Gennes-type free energy given by (assuming isotropic elastic stiffness $K$) 
\begin{align}
        F = \frac{1}{2}\int_{\mathbf{r}}\bigg\{ &A\left[(1-r)\mathrm{Tr}\mathbf{Q}^2+r(\mathrm{Tr}\mathbf{Q}^2)^2\right]+K(\partial_kQ_{ij})^2\notag\\
        &+ \kappa (\partial_i\partial_jQ_{kl})^2\bigg\}\;,
    \label{eq:Landau}
\end{align}
which corresponds to a molecular field given by 
\begin{equation}
\mathbf{H} = -\frac{\delta F}{\delta \mathbf{Q}}= -A(1-r+2rTr\Q^2)\Q+K\nabla^2\Q-\kappa\nabla^4\Q.
\end{equation}
Here, $A$ is a condensation energy and $r$ controls the passive transition between the nematic ($r>1$) and isotropic ($r<1$) phases. Below we focus on the behavior in the nematic phase,  where the equilibrium magnitude is $S_0=\sqrt{(r-1)/r}$. Finally, $\kappa$ represents  an effective surface tension {(see Appendix \ref{app:C})}, assumed isotropic for simplicity. As discussed below, this term provides stability at small length scales.

At the low Reynolds numbers relevant to active nematics, the flow field is determined by the Stokes equation
\begin{equation}
\Gamma \mathbf{u}=- \bm\nabla p + \bm\nabla \cdot \bm\sigma^a\;,
\label{eq:u}
\end{equation}
which balances frictional forces controlled by the drag $\Gamma$, the gradient of pressure $p$, and an active stress $\bm\sigma^a=\alpha\Q$. 
Here we focus on extensile systems because, to our knowledge, most of the existing realizations of active nematics are extensile. As described in  Ref.~\cite{giomi2010sheared}, to linear order the behavior of active nematics is controlled by the parameter $\alpha\lambda$: there is a duality between, for instance,  extensile ($\alpha<0$), rod-like and flow aligning ($\lambda>1$) fluids and contractile ($\alpha>0$), disk-shaped and flow tumbling  ($-1<\lambda<0$) ones. We have found that this similarity extends to the nonlinear behavior, and the phase diagram for contractile, flow-tumbling discotics has the same qualitative form as shown in Fig. 1. In active fluids with $\alpha\lambda<1$ the uniform state is linearly stable. 
We have neglected the elastic stress, whose effect on the linear stability of a uniform nematic state is captured by the phenomenological surface tension term with parameter $\kappa$ (see Appendices \ref{app:A} and \ref{app:C}). This term is needed to provide stability at short scale when the effective nematic stiffness becomes negative due to activity.

We consider dense active nematics with constant density in two situations: (i) an incompressible fluid with $\bm\nabla\cdot\bu=0$, and 
(ii) a fluid where the density is maintained constant, for instance by birth and death processes, without enforcing  $\bm\nabla\cdot\bu=0$. We refer to this second case as a compressible fluid. In the first case, the pressure serves as a Lagrange multiplier used to enforce the constraint of incompressibility. In the second case, the pressure is assumed to only depend on density, and therefore pressure gradients drop out of the force balance equation due to the constraint of constant density. This compressible, but constant density, limit has been used in previous literature as a minimal model of active flows on a substrate \cite{srivastava2016negative,oza2016antipolar,shankar2018defect}.  Additionally, as discussed further below, the behavior obtained is qualitatively similar to that obtained in particle simulations  of dry (i.e., friction-dominated) truly compressible fluids, where the density is allowed to vary \cite{decamp2015orientational,putzig2016instabilities, patelli2019understanding}. It therefore provides a useful minimal framework for describing compressible active flows on a substrate. Comparing the two models clarifies the role hydrodynamic effects, introduced by the incompressibility constraints, play in controlling emergent states in active nematics.

Our hydrodynamic model contains several length scales. The coherence length $\xi=\sqrt{K/(A(r-1))}$ controls the minimal scale for spatial variations of the order parameter. 
We work here deep in the nematic regime with $r=2$, giving an equilibrium order parameter $S_0=\sqrt{1/2}$. The active length $\ell_a=\sqrt{K/|\alpha|}$ quantifies the relative importance of active and elastic stresses.  Finally, the length scale $\ell_\kappa = \sqrt{\kappa/K}$ controls the typical size of smooth pattern formations in the $\Q$ tensor. We have chosen the nematic correlation length $\xi$ as the unit of length, the nematic relaxation time $\tau=\gamma/A$ as the unit of time and $\Gamma\xi^2/\tau$ as the unit of stress. 
In these units our equations only contain four dimensionless parameters: the flow alignment parameter which we fix at $\lambda=1.5$, $r=2$ that tunes the distance from the passive isotropic-nematic transition, $\tilde\kappa=\kappa/(K\xi^2)=(\ell_\kappa/\xi)^2$ which we choose to be $1$ unless otherwise specified, and a dimensionless activity  $\tilde{\alpha}=\alpha\gamma/(\Gamma K)$. 
In the following we drop the tildes on the dimensionless variables.

We solve Eqs.~(\ref{eq:Qeq}) and (\ref{eq:u}) numerically using the finite difference method in a square periodic box of size $L=128$. Spatial derivatives are evaluated with a fourth-order central difference on a uniform square grid with spacing $h=0.5$. To integrate over time, we use the Euler scheme with a time step $\delta t=10^{-3}$. Finally, we use the vorticity-stream function formulation to account for flow incompressibility \cite{chung2010}. The values of the parameters are listed in Table \ref{tab:paras}.

\begin{table}[h]
\centering
\caption{{\label{tab:paras} Values of parameters.}}
\begin{tabular}{p{0.2\textwidth}p{0.15\textwidth}p{0.1\textwidth}}  \hline
unit length &unit time &unit stress\\   $\xi=\sqrt{K/A(r-1)}$ &$\tau=\gamma/A$ &$\Gamma\xi^2/\tau$ \\ \hline
$r$ &$\lambda$ &$\kappa$ (default)\\   
2 &1.5 &1 \\ \hline
$L$ &$h$ &$\delta t$ \\  
128 &0.5 &0.001 \\ \hline
\end{tabular}
\end{table}

\section{Numerical State diagram}
\label{sec:numerics}

\begin{figure*}[t]
	\includegraphics[angle=0,width=1\textwidth]{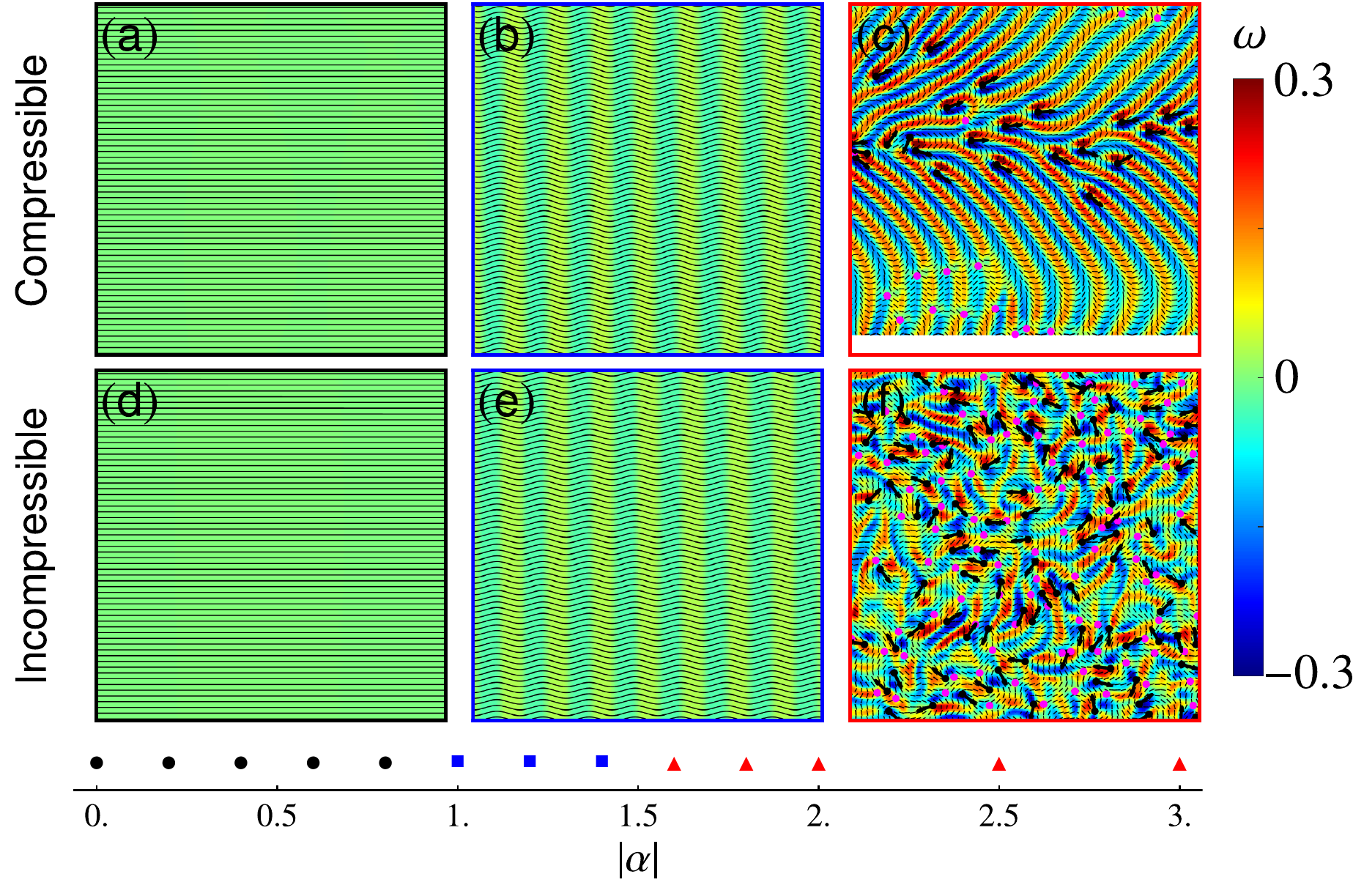}\\
\caption{\label{fig:PhasDiag} State diagram for (a-c) compressible and (d-f) incompressible systems. The background color in each snapshot shows the vorticity $\omega$ with magnitude corresponding to the colorbar on the far right. {The black arrows are $+1/2$ defects, whose heads indicate the defect polarization. Magenta dots are $-1/2$ defects.}  All simulations start from a uniform director oriented along $x$ with $S=\sqrt{1/2}$. Some small noise is added to the orientation to initiate the instability. For both compressible and incompressible systems, the uniform nematic state is stable at $|\alpha|<1$ (black dots, panels a and d). Then, at $1<|\alpha|<1.5$, the uniform state becomes unstable and both systems develop a parallel band state through the bending instability typically found in active nematics (blue squares, panels b and e). At $|\alpha|>1.5$, the band state can no longer accommodate the strong activity. In this case, defect flocking emerges in compressible systems while dynamics in incompressible systems becomes chaotic (red triangles, panels c and f).
}
\label{fig:PD} 
\end{figure*}

Figure \ref{fig:PD} shows typical configurations of active nematics in compressible versus incompressible flows, starting from a homogeneous ordered state and increasing activity. At low activity, the homogeneous state remains stable, for both compressible and incompressible systems (Figs. \ref{fig:PD}a and \ref{fig:PD}d), due to the presence of substrate friction.
Increasing activity destabilizes the homogeneous state, giving rise to a bend pattern {(for $\lambda>1$)} in the director field (Figs. \ref{fig:PD}b, \ref{fig:PD}e, {and Movies 1 and 2}).
The instability threshold is the same for both compressible and incompressible systems and is consistent with the value obtained from linear stability analysis \cite{srivastava2016negative}, which predicts a critical activity $|\alpha_{c0}|=\frac{2K\Gamma}{\gamma(\lambda+S_0)}\approx0.91$.
At higher activity, however, relaxing the incompressibility constraint changes fundamentally the character of the defect patterns. Incompressible systems transition  to a state of defect chaos (Fig. \ref{fig:PD}f {and Movie 4}), with proliferation of half-integer defects in the nematic texture and spatio-temporal chaotic vortical flows, as observed in many previous studies \cite{giomi2014defect, doostmohammadi2018active}. When the flow is compressible defect pairs also unbind at high activity. However, instead of the familiar chaotic dynamics, we observe a well-organized structure, consisting of a smectic-like state of equally spaced N\'eel or kink walls that have been referred to as arches due to the structure of the director field (Fig.~\ref{fig:PD}c) ~\cite{patelli2019understanding}, with associated polar order of $+1/2$ defects. Upon nucleation of defect pairs, the  $+1/2$  defects move away from their $-1/2$ companions, leaving a kink-wall structure of the director field in their wake. Active torques align the defect and the kink walls, resulting in  polar flocking of the $+1/2$ defects. This ordered state of defect and nematic texture was found to be a stable solution of the hydrodynamic of a defect gas derived in Ref.~\cite{shankar2019hydrodynamics}.  These structures have been previously found to be stable in continuum simulation of dry active nematics~\cite{decamp2015orientational,putzig2016instabilities}, but seemed to be long-lived metastable states in particle simulations of the same system~\cite{patelli2019understanding}.
They are also similar to the filamentous network of domain walls that was recently shown to dominate the coarsening dynamics of polar active matter~\cite{chardac2021topology}. In our compressible dry nematic, however, arches form a stable ordered state.

\section{Arch patterns and stability}
\label{sec:KW}
\begin{figure}[t!]
	\includegraphics[angle=0,width=0.98\columnwidth]{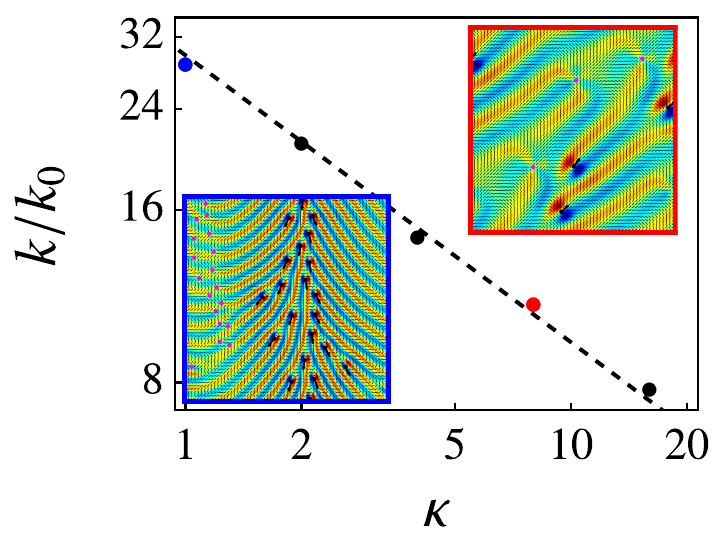}\\
\caption{Wavenumber $k$ corresponding to the peak of the Fourier transform of the nematic order phase field in the arch state as a function of $\kappa$ on a log-log scale. $k$ is scaled by $k_0=\pi/L$, which is the minimal wavenumber allowed by the finite system size. The two insets show typical snapshots from simulations at $\kappa=1$ (blue box, corresponding to the blue dot on the plot) and $\kappa=8$ (red box, {corresponding to the red dot on the plot}), respectively, and $\alpha=-2$. 
}
\label{fig:KW} 
\end{figure}

To quantify the periodicity of arches, we Fourier transform the phase field $\theta(\mathbf{r})$ obtained from numerical simulations in the regime where arches coexist with topological defects. The peak of the Fourier spectrum allows us to extract a characteristic wavenumber $k$ of the arch pattern, and the results show that $k\sim \ell_\kappa^{-1}$, where $\ell_\kappa = \sqrt{\kappa/K}$. Figure~\ref{fig:KW} shows $k$ as a function $\kappa$ on a log-log scale at a fixed activity $\alpha=-2$, demonstrating that indeed $k\sim\kappa^{-1/2}$.

\begin{figure}[t]	
\includegraphics[angle=0,width=0.98\columnwidth]{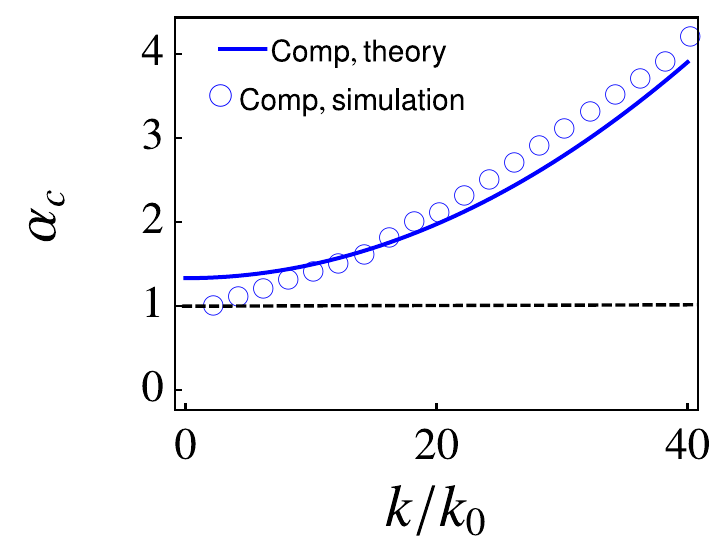}
\caption{Critical activity above which a solution of uniform arches of arch wavenumber $k$ becomes linearly unstable for a compressible system. The blue circles are the values obtained from numerical simulations, while the blue solid line shows the prediction the analytical linear stability analysis in Eq.~ (\ref{eq:alphac}). The black dashed line corresponds to the critical activity above which the uniform state ($k=0$) becomes unstable. The numerical parameter values are listed in Table~\ref{tab:paras}.}
\label{fig:alphac_KW} 
\end{figure}

To examine the stability of uniform arches in compressible flows, we solve numerically Eqs. (\ref{eq:Qeq}) and (\ref{eq:u}) with an initial condition given by an arbitrary number $n$ of perfectly aligned arches. Specifically, we set $S=\sqrt{(r-1)/r}$ and impose an initial texture $\theta(\mathbf{r})=n\pi x/L$ (mod $\pi$), with {$n$} an integer representing the number of equilibrium arches that can be accommodated in a system of size $L$.
Figure~\ref{fig:alphac_KW} shows the critical activity $|\alpha_c|$ as function of the number of initial arches $n$ and above which the initial uniform arch pattern becomes unstable to defect nucleation.
We notice that $|\alpha_c|$ increases monotonically with the initial arch wavenumber $k$ implying that the transition to the regime of coexistence between arches and orientational defects depends on initial conditions, thus the metastability of arches. The numerical protocol of destabilizing initially uniform arches is done in the absence of any noise and by gradually increasing activity. In this case, arches remain metastable also for $|\alpha|<1$. However, any additional noise perturbations will melt the arches into the uniform ordered state as shown in the state diagram in Fig.~(\ref{fig:PhasDiag}). This is not the case for $|\alpha_c|>1$ where many possible arch states can be stable as discussed next.

By a theoretical stability analysis, we can also further understand the role of flow compressibility in stabilizing arch patterns. To do this, it is convenient to write Eqs.~ (\ref{eq:Qeq}) and (\ref{eq:u}) in complex coordinates by introducing $z=x+iy$, $\partial_z=(\partial_x - i\partial_y)/2$, $\partial_{\bar{z}}=(\partial_x + i\partial_y)/2$, and $\partial^2=\partial_z \partial_{\bar{z}}$. The Q-tensor and flow velocity are written equivalently as complex scalar fields, $\psi = S\exp\{2i\theta\}$ and $u=u_x + i u_y$. In this notation, Eqs.~(\ref{eq:Qeq}) and (\ref{eq:u}) take the form

\begin{align}
\partial_t \psi=&-u\partial_z \psi-\bar{u}\partial_{\bar{z}}\psi+(\partial_z u - \partial_{\bar{z}}\bar{u})
\psi + \lambda \partial_{\bar{z}}u \nonumber\\
&- \frac{A}{\gamma}(1-r+4r|\psi|^2)\psi+\frac{4K}{\gamma}\partial^2\psi-\frac{16\kappa}{\gamma}\partial^2\partial^2\psi\;,
\label{eq:psi}
\end{align}
and
\begin{equation}
\Gamma u=2\alpha\partial_z\psi-2\partial_{\bar{z}} p\;.
\label{eq:u_cmplx}
\end{equation}

The incompressibility constraint reads $\partial_{\bar{z}}\bar{u}+\partial_{z}u=0$.
The first two terms on the right hand side of Eq.~(\ref{eq:psi}) are the convective terms, which are followed by the terms for vorticity.

When the flow is compressible, the pressure gradients vanishes and the flow velocity can be eliminated from the $\psi$ evolution. We seek a steady-state solution for perfectly aligned arches 
\begin{equation}
\psi_k^0(x) = S_k \exp\{2i kx\}\;,
\label{eq:arch_solution}
\end{equation}
where $S_k$ is a complex amplitude. In this case, the advective term and the term coupling to vorticity cancel each other, such that we obtain an equation for $S_k$ as 
\begin{equation}
\begin{aligned}
\partial_t S_k=&
-\frac{2\alpha\lambda}{\Gamma}k^2 S_k-\frac{A}{\gamma}(1-r+r|S_k|^2)S_k\\
&-\frac{4K}{\gamma}k^2 S_k-\frac{16\kappa}{\gamma}k^4 S_k\;.
\end{aligned}
\label{eq:psi_arch}
\end{equation}
This equation has a nontrivial steady-state solution given by 
\begin{align}
S_k =\sqrt{\frac{r-1-\epsilon_k}{r}}\;
 \label{eq:arch}
\end{align}
 where 
\begin{align}
 \epsilon_k=\left(\frac{4K}{A}+\frac{2\alpha\lambda\gamma}{\Gamma A}+\frac{16\kappa}{A}k^2\right)k^2\;.
\end{align}
For $k=0$, $\epsilon_k$ vanishes and this reduces to the homogeneous nematics.

To examine the stability of this arch solution against linear perturbations, we let 
\begin{equation}
\psi(z,\bar z) = \psi_k^0(x)+\delta\psi(z,\bar z)
\end{equation}
and linearize Eqs. (\ref{eq:Qeq}) and (\ref{eq:u}) in $\delta\psi$, leading to
\begin{equation}
\begin{aligned}
\partial_t \delta \psi=&- \frac{2\alpha ik}{\Gamma}(2 \psi_k^0\partial_z\delta \psi + \psi_k^0\partial_{\bar z}\overline{\delta \psi}-\overline{\psi_k^0}\partial_{\bar z}\delta \psi)\\
-&\frac{2\alpha}{\Gamma}k^2( \psi_k^0-\overline{\psi_k^0})\delta \psi+\frac{2\alpha}{\Gamma}\psi_k^0(\partial_z^2\delta \psi-\partial_{\bar z}^2\overline{\delta \psi})\\
+&(\frac{2\alpha\lambda}{\Gamma}+\frac{4K}{\gamma})\partial^2\delta \psi-\frac{16\kappa}{\gamma}\partial^2\partial^2\delta \psi\\
-&\frac{A}{\gamma}(1-r+rS_k^2)\delta \psi-\frac{A}{\gamma}4r \psi_k^0(\overline{\psi_k^0}\delta \psi+ \psi_k^0\overline{\delta \psi})
\label{eq:linear}\;.
\end{aligned}
\end{equation}
Since Eq.~(\ref{eq:linear}) is a linear differential equation with $x$-dependent coefficients, Fourier modes with different wavenumbers along $x$ are generally coupled and the eigenfunctions are not plane waves. One approach would then be to truncate the Fourier expansion to some order and numerically diagonalize the operator on the right hand side of Eq.~(\ref{eq:linear}),  as done for instance in Ref.~\cite{jiang2019trait}. For the purpose of determining the critical activity, we find it suffices, however, to retain only the wavenumber $k$ that sets the periodicity of the arch solution and write
\begin{equation}
\delta\psi(x,y) =  e^{2ikx}\sum_q \delta\psi_k(q) e^{iqy}\;.
\label{eq:del_Q}
\end{equation}

Substituting Eq.~(\ref{eq:del_Q}) into Eq.~(\ref{eq:linear}) and keeping only the lowest order mode in $k$, we find that, given $S_k^2$ is real, the $4\times 4$ dynamical equations for $\delta\psi_k(q)$, $\delta\psi_k(-q)$,  $\overline{\delta\psi_k(q)}$ and $\overline{\delta\psi_k(-q)}$ is block diagonal. The problem of finding the relaxation rate then reduces to the solution of two coupled equations that can be written as

\begin{equation}
\partial_t \begin{pmatrix}
\delta \psi_k(q)\\ \overline{\delta \psi_k(-q)}
\end{pmatrix}=
\begin{pmatrix}
m_{1} & m_{2}\\
m_{2} & m_{1}
\end{pmatrix} \begin{pmatrix}
\delta \psi_k(q)\\ \overline{\delta \psi_k(-q)}
\end{pmatrix}\;,
\label{eq:matrix}
\end{equation}

where 
\begin{align}
m_{1}=&
-\left(\frac{K}{\gamma}+\frac{\alpha\lambda}{2\Gamma}+\frac{8\kappa}{\gamma}k^2\right)q^2-\frac{\kappa}{\gamma}q^4\notag\\
&-\frac{A}{\gamma}(r-1)+\left(\frac{4K}{\gamma}+\frac{2\alpha\lambda}{\Gamma}+\frac{16\kappa}{\gamma}k^2\right)k^2\notag\\
m_{2}=&
-\frac{A}{\gamma}(r-1)+\left(\frac{4K}{\gamma}+\frac{2\alpha\lambda}{\Gamma}+\frac{16\kappa}{\gamma}k^2\right)k^2 \label{eq:matrix_elements}\;.
\end{align}

The eigenvalues of the dynamical matrix are $\nu_\pm=m_1\pm m_2$ and correspond to the growth rates of the two modes. The instability is controlled by the largest growth rate $\nu_-$,  given by
\begin{equation}
\nu_-(q)=  -\left(\frac{K}{\gamma}+\frac{\alpha\lambda}{2\Gamma}+\frac{8\kappa}{\gamma}k^2\right)q^2-\frac{\kappa}{\gamma}q^4\;.  
\end{equation}
Arches are therefore unstable for magnitudes of activity larger than 
the critical value
\begin{equation}
\alpha_c(k)=\frac{2\Gamma}{\lambda\gamma}\left(K+ 8\kappa k^2\right)\;.
\label{eq:alphac}
\end{equation}
where the coefficient of $q^2$ changes sign.
The wavenumber of the most rapidly growing mode $q_c=\sqrt{\frac{\gamma\lambda}{4\kappa\Gamma}}~(|\alpha|-\alpha_c)^{1/2}$ vanishes at the onset of instability.

The analytically-predicted dependence of $\alpha_c$ on the  arches wavenumber $k$ is shown as a solid blue line in Fig. \ref{fig:alphac_KW} and compares well with the results of numerical simulations.



\section{Defect ordering}
\label{sec:order}
\begin{figure}[h!]
	\includegraphics[angle=0,width=1\columnwidth]{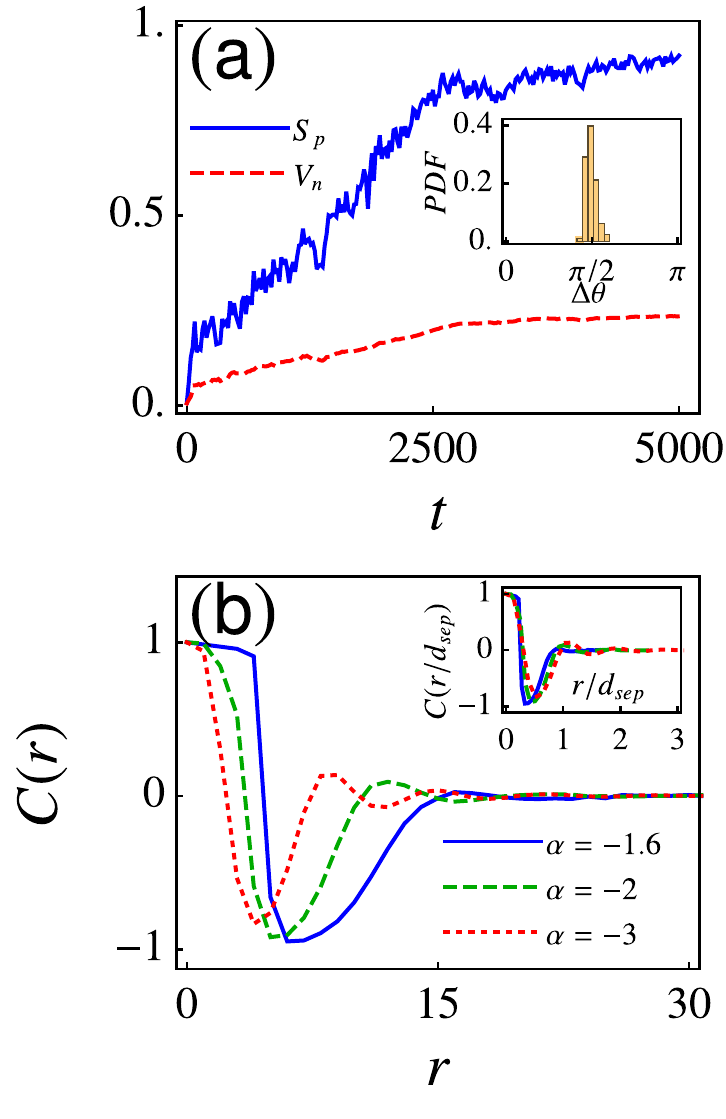}
\caption{\label{fig:VnSp} Defect order: {(a) For compressible systems the mean defect polarization $S_{p}$ (solid blue) and mean phase gradient $V_{n}$ (dashed red). Both grow with time and eventually saturate roughly at the same time.} The inset shows the probability distribution of the angle between $S_{p}$ and $V_{n}$, with a strong peak at $\pi/2$ (activity $\alpha = -2$). {(b) For incompressible systems, in contrast, there is only local (nematic) order and no long-range order. This is evident from the behavior of the correlation function of the defect polarization $C(r)$ for three different activities.} The inset shows the same correlation function with distance $r$ rescaled by the mean defect separation, $d_{sep}$.
}
\end{figure}

To quantify orientational order of defects, we define the polarization of 
the $i$-th $+1/2$ defect as  $\mathbf{p}_i=(\nabla\cdot \Q/|\nabla\cdot \Q|)\Bigr|_{\substack{\mathbf{r}=\mathbf{r}_i}}$ \cite{vromans2016orientational, tang2017orientation}.
In the compressible fluid, we measure a vector global polar order parameter of the $+1/2$ defects as an ensemble average of the defect polarizations given by
\begin{equation}
    \textbf{S}_p(t)=\frac{1}{N}\sum_{i=1}^N\textbf{p}_i(t)\;,
\end{equation} 
where $N$ is the number of $+1/2$ defects in the system at time $t$. 
To quantify the correlation between arches and polar order of $+1/2$ defects, we additionally measure the mean (spatially-averaged) phase gradient
\begin{equation}
    \mathbf{V}_n(t) = \langle \bm\nabla \theta(t)\rangle\;,
\end{equation} 
{as a global indicator of the arches periodicity.} Figure \ref{fig:VnSp}(a) shows that the magnitudes of both these  measures increase in time in a correlated manner, and both saturates to a constant value on similar time scales. The inset is the histogram of the angle between $\textbf{S}_p(t)$ and $\mathbf{V}_n(t)$ and is strongly peaked at $\pi/2$, demonstrating that the polar defect order is normal to the arches periodicity. These findings are consistent with the hydrodynamic theory presented in Ref.~ \cite{shankar2019hydrodynamics}.
In an incompressible fluid, defects do not reveal  global order, but may exhibit local order, as previously reported in Ref.~\cite{pearce2021orientational}, and which is extracted from the radial dependence of the defect orientation correlation function on the distance $r$ between defects
\begin{equation}
    C(r)= \frac{1}{N_{r}}\sum_{r<|\textbf{r}_i-\textbf{r}_j|<r+\delta r} \textbf{p}_i\cdot\textbf{p}_j\;,
\end{equation} 
where $N_r$ is the total number of defect pairs with separation within $(r,r+\delta r)$. Figure \ref{fig:VnSp}(b) shows the correlation function for various values of activity $\alpha$.
$C(r)$ has a pronounced dip at small distances, signifying local nematic defect order. Furthermore, when the distance $r$ is scaled by the average defect separation $d_{sep}$,  $C(r)$ for different values of activity nearly collapse into a single curve (inset of Fig. \ref{fig:VnSp}(b)).
Our findings on  local nematic defect order are  consistent with those reported in Ref. \cite{pearce2021orientational} for an incompressible active fluid with no substrate friction.

\section{Conclusions}
\label{sec:conclusion}
We have examined numerically and analytically the emergent dynamics of  two-dimensional active nematics on a substrate with compressible and  incompressible fluid flow. We find that long-range constraints imposed by incompressibility has a profound influence of the spontaneous flow structures observed upon increasing activity. While incompressible systems transition from a stationary homogeneous state to chaotic spatiotemporal dynamics with proliferation of unbound nematic defect pairs, compressible ones organize in a smectic-like state of aligned arches in the nematic texture and associated polar order of $+1/2$ defects. We demonstrate explictely that arch patterns are stable in the compressible case below an arch-width dependent critical activity, which we calculate analytically.  These arches underlie defect ordered states previously reported in compressible fluids that incorporate density fluctuations~\cite{putzig2016instabilities,patelli2019understanding}.

To put our results in context, we briefly compare them to previous work. 
Polar order of $+1/2$ defect and associated smectic arrangement of arches has been reported previously in simulations of hard spherocylinders~\cite{decamp2015orientational} and in numerical solution of the continuum hydrodynamics of dry active nematic both in the ``compressible’’ limit implemented here~\cite{srivastava2016negative}, as well as for a truly incompressible fluid with conserved density allowed to fluctuate~\cite{putzig2016instabilities}. 
Oza and Dunkel~\cite{oza2016antipolar} also used a dry continuum model, but placed themselves directly in the unstable regime and neglected entirely the rotational effects of the flow. In this limit they observed a variety of ordered structures, including defect lattices and anti-polar order of defects. Due to these differences in the model, a direct comparison with other results is not straightforward. Arches and associated polar defect order have also been observed in numerical simulations of Viscek-like model of self-propelled point particles that periodically reverse their direction of motion and interact through both aligning and repulsive interactions~\cite{patelli2019understanding}. Arches solutions where, however, found to be metastable in an associated continuum formulation of the same model, and no defect order was obtained in the continuum model.

Simulations of incompressible wet active nematics where dissipation is controlled entirely by viscous stresses have revealed local antialignment of $+1/2$ defects, but no longer range order~\cite{pearce2021orientational}, as we find here in the dry limit. Finally, Nejad \textit{et al.}~\cite{nejad2021memory} have recently examined the interplay between viscous and frictional dissipation in incompressible fluids. At large friction they observe arches and local polar defect order, but this behavior seems to be transient. In fact we have also observed such structures in the incompressible case, but they are never present in the long-time steady state, consistent with our analytical results on arches solutions. A more detailed comparison with their work is, however, not possible since these authors use the shear viscosity, which is zero in our work, to determine the units of time. Additionally, their large friction limit can also be interpreted as the limit of very low activity. It is therefore not surprising that in these limit they find no defect proliferation.

Our work demonstrates that  nematic texture and defect order in active liquid crystals depend strongly on the nature of the flows. It further identifies the origin of apparent discrepancies reported in the literature on polar or nematic order of defects as arising from the constraint imposed by incompressibility.

Finally, it may be possible to observe the smectic arch state and polar defect order in dense, possibly jammed active systems where friction is the primary dissipation mechanism. In fact experiments in dense monolayer of spindle-shaped cells have revealed structures that resemble arches at high cell density, where the the cells are essentially jammed and defects are no longer motile~\cite{duclos2017topological}. The phenomena described here may also be relevant to bacterial colonies growing on a frictional substrate.
\\

\begin{acknowledgments}
This work was supported by the National Science Foundation Grants No.~DMR-2041459 (Z.Y. and M.C.M.)
and No.~PHY-1748958 (S.P, Z.C. and M.J.B.). 
\end{acknowledgments}

\bibliography{refs}

\begin{thebibliography}{36}%
\makeatletter
\providecommand \@ifxundefined [1]{%
 \@ifx{#1\undefined}
}%
\providecommand \@ifnum [1]{%
 \ifnum #1\expandafter \@firstoftwo
 \else \expandafter \@secondoftwo
 \fi
}%
\providecommand \@ifx [1]{%
 \ifx #1\expandafter \@firstoftwo
 \else \expandafter \@secondoftwo
 \fi
}%
\providecommand \natexlab [1]{#1}%
\providecommand \enquote  [1]{``#1''}%
\providecommand \bibnamefont  [1]{#1}%
\providecommand \bibfnamefont [1]{#1}%
\providecommand \citenamefont [1]{#1}%
\providecommand \href@noop [0]{\@secondoftwo}%
\providecommand \href [0]{\begingroup \@sanitize@url \@href}%
\providecommand \@href[1]{\@@startlink{#1}\@@href}%
\providecommand \@@href[1]{\endgroup#1\@@endlink}%
\providecommand \@sanitize@url [0]{\catcode `\\12\catcode `\$12\catcode
  `\&12\catcode `\#12\catcode `\^12\catcode `\_12\catcode `\%12\relax}%
\providecommand \@@startlink[1]{}%
\providecommand \@@endlink[0]{}%
\providecommand \url  [0]{\begingroup\@sanitize@url \@url }%
\providecommand \@url [1]{\endgroup\@href {#1}{\urlprefix }}%
\providecommand \urlprefix  [0]{URL }%
\providecommand \Eprint [0]{\href }%
\providecommand \doibase [0]{http://dx.doi.org/}%
\providecommand \selectlanguage [0]{\@gobble}%
\providecommand \bibinfo  [0]{\@secondoftwo}%
\providecommand \bibfield  [0]{\@secondoftwo}%
\providecommand \translation [1]{[#1]}%
\providecommand \BibitemOpen [0]{}%
\providecommand \bibitemStop [0]{}%
\providecommand \bibitemNoStop [0]{.\EOS\space}%
\providecommand \EOS [0]{\spacefactor3000\relax}%
\providecommand \BibitemShut  [1]{\csname bibitem#1\endcsname}%
\let\auto@bib@innerbib\@empty
\bibitem [{\citenamefont {Doostmohammadi}\ \emph {et~al.}(2018)\citenamefont
  {Doostmohammadi}, \citenamefont {Ign{\'e}s-Mullol}, \citenamefont {Yeomans},\
  and\ \citenamefont {Sagu{\'e}s}}]{doostmohammadi2018active}%
  \BibitemOpen
  \bibfield  {author} {\bibinfo {author} {\bibfnamefont {A.}~\bibnamefont
  {Doostmohammadi}}, \bibinfo {author} {\bibfnamefont {J.}~\bibnamefont
  {Ign{\'e}s-Mullol}}, \bibinfo {author} {\bibfnamefont {J.~M.}\ \bibnamefont
  {Yeomans}}, \ and\ \bibinfo {author} {\bibfnamefont {F.}~\bibnamefont
  {Sagu{\'e}s}},\ }\href@noop {} {\bibfield  {journal} {\bibinfo  {journal}
  {Nature communications}\ }\textbf {\bibinfo {volume} {9}},\ \bibinfo {pages}
  {1} (\bibinfo {year} {2018})}\BibitemShut {NoStop}%
\bibitem [{\citenamefont {Sanchez}\ \emph {et~al.}(2012)\citenamefont
  {Sanchez}, \citenamefont {Chen}, \citenamefont {DeCamp}, \citenamefont
  {Heymann},\ and\ \citenamefont {Dogic}}]{sanchez2012spontaneous}%
  \BibitemOpen
  \bibfield  {author} {\bibinfo {author} {\bibfnamefont {T.}~\bibnamefont
  {Sanchez}}, \bibinfo {author} {\bibfnamefont {D.~T.}\ \bibnamefont {Chen}},
  \bibinfo {author} {\bibfnamefont {S.~J.}\ \bibnamefont {DeCamp}}, \bibinfo
  {author} {\bibfnamefont {M.}~\bibnamefont {Heymann}}, \ and\ \bibinfo
  {author} {\bibfnamefont {Z.}~\bibnamefont {Dogic}},\ }\href@noop {}
  {\bibfield  {journal} {\bibinfo  {journal} {Nature}\ }\textbf {\bibinfo
  {volume} {491}},\ \bibinfo {pages} {431} (\bibinfo {year}
  {2012})}\BibitemShut {NoStop}%
\bibitem [{\citenamefont {Zhang}\ \emph {et~al.}(2018)\citenamefont {Zhang},
  \citenamefont {Kumar}, \citenamefont {Ross}, \citenamefont {Gardel},\ and\
  \citenamefont {De~Pablo}}]{zhang2018interplay}%
  \BibitemOpen
  \bibfield  {author} {\bibinfo {author} {\bibfnamefont {R.}~\bibnamefont
  {Zhang}}, \bibinfo {author} {\bibfnamefont {N.}~\bibnamefont {Kumar}},
  \bibinfo {author} {\bibfnamefont {J.~L.}\ \bibnamefont {Ross}}, \bibinfo
  {author} {\bibfnamefont {M.~L.}\ \bibnamefont {Gardel}}, \ and\ \bibinfo
  {author} {\bibfnamefont {J.~J.}\ \bibnamefont {De~Pablo}},\ }\href@noop {}
  {\bibfield  {journal} {\bibinfo  {journal} {Proceedings of the National
  Academy of Sciences}\ }\textbf {\bibinfo {volume} {115}},\ \bibinfo {pages}
  {E124} (\bibinfo {year} {2018})}\BibitemShut {NoStop}%
\bibitem [{\citenamefont {Thutupalli}\ \emph {et~al.}(2015)\citenamefont
  {Thutupalli}, \citenamefont {Sun}, \citenamefont {Bunyak}, \citenamefont
  {Palaniappan},\ and\ \citenamefont {Shaevitz}}]{thutupalli2015directional}%
  \BibitemOpen
  \bibfield  {author} {\bibinfo {author} {\bibfnamefont {S.}~\bibnamefont
  {Thutupalli}}, \bibinfo {author} {\bibfnamefont {M.}~\bibnamefont {Sun}},
  \bibinfo {author} {\bibfnamefont {F.}~\bibnamefont {Bunyak}}, \bibinfo
  {author} {\bibfnamefont {K.}~\bibnamefont {Palaniappan}}, \ and\ \bibinfo
  {author} {\bibfnamefont {J.~W.}\ \bibnamefont {Shaevitz}},\ }\href@noop {}
  {\bibfield  {journal} {\bibinfo  {journal} {Journal of The Royal Society
  Interface}\ }\textbf {\bibinfo {volume} {12}},\ \bibinfo {pages} {20150049}
  (\bibinfo {year} {2015})}\BibitemShut {NoStop}%
\bibitem [{\citenamefont {Doostmohammadi}\ \emph
  {et~al.}(2016{\natexlab{a}})\citenamefont {Doostmohammadi}, \citenamefont
  {Thampi},\ and\ \citenamefont {Yeomans}}]{doostmohammadi2016defect}%
  \BibitemOpen
  \bibfield  {author} {\bibinfo {author} {\bibfnamefont {A.}~\bibnamefont
  {Doostmohammadi}}, \bibinfo {author} {\bibfnamefont {S.~P.}\ \bibnamefont
  {Thampi}}, \ and\ \bibinfo {author} {\bibfnamefont {J.~M.}\ \bibnamefont
  {Yeomans}},\ }\href@noop {} {\bibfield  {journal} {\bibinfo  {journal}
  {Physical review letters}\ }\textbf {\bibinfo {volume} {117}},\ \bibinfo
  {pages} {048102} (\bibinfo {year} {2016}{\natexlab{a}})}\BibitemShut
  {NoStop}%
\bibitem [{\citenamefont {Duclos}\ \emph {et~al.}(2017)\citenamefont {Duclos},
  \citenamefont {Erlenk{\"a}mper}, \citenamefont {Joanny},\ and\ \citenamefont
  {Silberzan}}]{duclos2017topological}%
  \BibitemOpen
  \bibfield  {author} {\bibinfo {author} {\bibfnamefont {G.}~\bibnamefont
  {Duclos}}, \bibinfo {author} {\bibfnamefont {C.}~\bibnamefont
  {Erlenk{\"a}mper}}, \bibinfo {author} {\bibfnamefont {J.-F.}\ \bibnamefont
  {Joanny}}, \ and\ \bibinfo {author} {\bibfnamefont {P.}~\bibnamefont
  {Silberzan}},\ }\href@noop {} {\bibfield  {journal} {\bibinfo  {journal}
  {Nature Physics}\ }\textbf {\bibinfo {volume} {13}},\ \bibinfo {pages} {58}
  (\bibinfo {year} {2017})}\BibitemShut {NoStop}%
\bibitem [{\citenamefont {Blanch-Mercader}\ \emph {et~al.}(2018)\citenamefont
  {Blanch-Mercader}, \citenamefont {Yashunsky}, \citenamefont {Garcia},
  \citenamefont {Duclos}, \citenamefont {Giomi},\ and\ \citenamefont
  {Silberzan}}]{blanch2018turbulent}%
  \BibitemOpen
  \bibfield  {author} {\bibinfo {author} {\bibfnamefont {C.}~\bibnamefont
  {Blanch-Mercader}}, \bibinfo {author} {\bibfnamefont {V.}~\bibnamefont
  {Yashunsky}}, \bibinfo {author} {\bibfnamefont {S.}~\bibnamefont {Garcia}},
  \bibinfo {author} {\bibfnamefont {G.}~\bibnamefont {Duclos}}, \bibinfo
  {author} {\bibfnamefont {L.}~\bibnamefont {Giomi}}, \ and\ \bibinfo {author}
  {\bibfnamefont {P.}~\bibnamefont {Silberzan}},\ }\href@noop {} {\bibfield
  {journal} {\bibinfo  {journal} {Physical review letters}\ }\textbf {\bibinfo
  {volume} {120}},\ \bibinfo {pages} {208101} (\bibinfo {year}
  {2018})}\BibitemShut {NoStop}%
\bibitem [{\citenamefont {Alert}\ \emph {et~al.}(2022)\citenamefont {Alert},
  \citenamefont {Casademunt},\ and\ \citenamefont {Joanny}}]{alert2022active}%
  \BibitemOpen
  \bibfield  {author} {\bibinfo {author} {\bibfnamefont {R.}~\bibnamefont
  {Alert}}, \bibinfo {author} {\bibfnamefont {J.}~\bibnamefont {Casademunt}}, \
  and\ \bibinfo {author} {\bibfnamefont {J.-F.}\ \bibnamefont {Joanny}},\
  }\href@noop {} {\bibfield  {journal} {\bibinfo  {journal} {Annual Review of
  Condensed Matter Physics}\ }\textbf {\bibinfo {volume} {13}} (\bibinfo {year}
  {2022})}\BibitemShut {NoStop}%
\bibitem [{\citenamefont {Giomi}\ \emph {et~al.}(2013)\citenamefont {Giomi},
  \citenamefont {Bowick}, \citenamefont {Ma},\ and\ \citenamefont
  {Marchetti}}]{giomi2013defect}%
  \BibitemOpen
  \bibfield  {author} {\bibinfo {author} {\bibfnamefont {L.}~\bibnamefont
  {Giomi}}, \bibinfo {author} {\bibfnamefont {M.~J.}\ \bibnamefont {Bowick}},
  \bibinfo {author} {\bibfnamefont {X.}~\bibnamefont {Ma}}, \ and\ \bibinfo
  {author} {\bibfnamefont {M.~C.}\ \bibnamefont {Marchetti}},\ }\href@noop {}
  {\bibfield  {journal} {\bibinfo  {journal} {Physical review letters}\
  }\textbf {\bibinfo {volume} {110}},\ \bibinfo {pages} {228101} (\bibinfo
  {year} {2013})}\BibitemShut {NoStop}%
\bibitem [{\citenamefont {Pismen}(2013)}]{pismen2013dynamics}%
  \BibitemOpen
  \bibfield  {author} {\bibinfo {author} {\bibfnamefont {L.~M.}\ \bibnamefont
  {Pismen}},\ }\href@noop {} {\bibfield  {journal} {\bibinfo  {journal}
  {Physical Review E}\ }\textbf {\bibinfo {volume} {88}},\ \bibinfo {pages}
  {050502(R)} (\bibinfo {year} {2013})}\BibitemShut {NoStop}%
\bibitem [{\citenamefont {Giomi}\ \emph {et~al.}(2014)\citenamefont {Giomi},
  \citenamefont {Bowick}, \citenamefont {Mishra}, \citenamefont {Sknepnek},\
  and\ \citenamefont {Cristina~Marchetti}}]{giomi2014defect}%
  \BibitemOpen
  \bibfield  {author} {\bibinfo {author} {\bibfnamefont {L.}~\bibnamefont
  {Giomi}}, \bibinfo {author} {\bibfnamefont {M.~J.}\ \bibnamefont {Bowick}},
  \bibinfo {author} {\bibfnamefont {P.}~\bibnamefont {Mishra}}, \bibinfo
  {author} {\bibfnamefont {R.}~\bibnamefont {Sknepnek}}, \ and\ \bibinfo
  {author} {\bibfnamefont {M.}~\bibnamefont {Cristina~Marchetti}},\ }\href@noop
  {} {\bibfield  {journal} {\bibinfo  {journal} {Philosophical Transactions of
  the Royal Society A: Mathematical, Physical and Engineering Sciences}\
  }\textbf {\bibinfo {volume} {372}},\ \bibinfo {pages} {20130365} (\bibinfo
  {year} {2014})}\BibitemShut {NoStop}%
\bibitem [{\citenamefont {Keber}\ \emph {et~al.}(2014)\citenamefont {Keber},
  \citenamefont {Loiseau}, \citenamefont {Sanchez}, \citenamefont {DeCamp},
  \citenamefont {Giomi}, \citenamefont {Bowick}, \citenamefont {Marchetti},
  \citenamefont {Dogic},\ and\ \citenamefont {Bausch}}]{keber2014topology}%
  \BibitemOpen
  \bibfield  {author} {\bibinfo {author} {\bibfnamefont {F.~C.}\ \bibnamefont
  {Keber}}, \bibinfo {author} {\bibfnamefont {E.}~\bibnamefont {Loiseau}},
  \bibinfo {author} {\bibfnamefont {T.}~\bibnamefont {Sanchez}}, \bibinfo
  {author} {\bibfnamefont {S.~J.}\ \bibnamefont {DeCamp}}, \bibinfo {author}
  {\bibfnamefont {L.}~\bibnamefont {Giomi}}, \bibinfo {author} {\bibfnamefont
  {M.~J.}\ \bibnamefont {Bowick}}, \bibinfo {author} {\bibfnamefont {M.~C.}\
  \bibnamefont {Marchetti}}, \bibinfo {author} {\bibfnamefont {Z.}~\bibnamefont
  {Dogic}}, \ and\ \bibinfo {author} {\bibfnamefont {A.~R.}\ \bibnamefont
  {Bausch}},\ }\href@noop {} {\bibfield  {journal} {\bibinfo  {journal}
  {Science}\ }\textbf {\bibinfo {volume} {345}},\ \bibinfo {pages} {1135}
  (\bibinfo {year} {2014})}\BibitemShut {NoStop}%
\bibitem [{\citenamefont {Kawaguchi}\ \emph {et~al.}(2017)\citenamefont
  {Kawaguchi}, \citenamefont {Kageyama},\ and\ \citenamefont
  {Sano}}]{kawaguchi2017topological}%
  \BibitemOpen
  \bibfield  {author} {\bibinfo {author} {\bibfnamefont {K.}~\bibnamefont
  {Kawaguchi}}, \bibinfo {author} {\bibfnamefont {R.}~\bibnamefont {Kageyama}},
  \ and\ \bibinfo {author} {\bibfnamefont {M.}~\bibnamefont {Sano}},\
  }\href@noop {} {\bibfield  {journal} {\bibinfo  {journal} {Nature}\ }\textbf
  {\bibinfo {volume} {545}},\ \bibinfo {pages} {327} (\bibinfo {year}
  {2017})}\BibitemShut {NoStop}%
\bibitem [{\citenamefont {Saw}\ \emph {et~al.}(2017)\citenamefont {Saw},
  \citenamefont {Doostmohammadi}, \citenamefont {Nier}, \citenamefont
  {Kocgozlu}, \citenamefont {Thampi}, \citenamefont {Toyama}, \citenamefont
  {Marcq}, \citenamefont {Lim}, \citenamefont {Yeomans},\ and\ \citenamefont
  {Ladoux}}]{saw2017topological}%
  \BibitemOpen
  \bibfield  {author} {\bibinfo {author} {\bibfnamefont {T.~B.}\ \bibnamefont
  {Saw}}, \bibinfo {author} {\bibfnamefont {A.}~\bibnamefont {Doostmohammadi}},
  \bibinfo {author} {\bibfnamefont {V.}~\bibnamefont {Nier}}, \bibinfo {author}
  {\bibfnamefont {L.}~\bibnamefont {Kocgozlu}}, \bibinfo {author}
  {\bibfnamefont {S.}~\bibnamefont {Thampi}}, \bibinfo {author} {\bibfnamefont
  {Y.}~\bibnamefont {Toyama}}, \bibinfo {author} {\bibfnamefont
  {P.}~\bibnamefont {Marcq}}, \bibinfo {author} {\bibfnamefont {C.~T.}\
  \bibnamefont {Lim}}, \bibinfo {author} {\bibfnamefont {J.~M.}\ \bibnamefont
  {Yeomans}}, \ and\ \bibinfo {author} {\bibfnamefont {B.}~\bibnamefont
  {Ladoux}},\ }\href@noop {} {\bibfield  {journal} {\bibinfo  {journal}
  {Nature}\ }\textbf {\bibinfo {volume} {544}},\ \bibinfo {pages} {212}
  (\bibinfo {year} {2017})}\BibitemShut {NoStop}%
\bibitem [{\citenamefont {Maroudas-Sacks}\ \emph {et~al.}(2021)\citenamefont
  {Maroudas-Sacks}, \citenamefont {Garion}, \citenamefont {Shani-Zerbib},
  \citenamefont {Livshits}, \citenamefont {Braun},\ and\ \citenamefont
  {Keren}}]{maroudas2021topological}%
  \BibitemOpen
  \bibfield  {author} {\bibinfo {author} {\bibfnamefont {Y.}~\bibnamefont
  {Maroudas-Sacks}}, \bibinfo {author} {\bibfnamefont {L.}~\bibnamefont
  {Garion}}, \bibinfo {author} {\bibfnamefont {L.}~\bibnamefont
  {Shani-Zerbib}}, \bibinfo {author} {\bibfnamefont {A.}~\bibnamefont
  {Livshits}}, \bibinfo {author} {\bibfnamefont {E.}~\bibnamefont {Braun}}, \
  and\ \bibinfo {author} {\bibfnamefont {K.}~\bibnamefont {Keren}},\
  }\href@noop {} {\bibfield  {journal} {\bibinfo  {journal} {Nature Physics}\
  }\textbf {\bibinfo {volume} {17}},\ \bibinfo {pages} {251} (\bibinfo {year}
  {2021})}\BibitemShut {NoStop}%
\bibitem [{\citenamefont {Copenhagen}\ \emph {et~al.}(2021)\citenamefont
  {Copenhagen}, \citenamefont {Alert}, \citenamefont {Wingreen},\ and\
  \citenamefont {Shaevitz}}]{copenhagen2021topological}%
  \BibitemOpen
  \bibfield  {author} {\bibinfo {author} {\bibfnamefont {K.}~\bibnamefont
  {Copenhagen}}, \bibinfo {author} {\bibfnamefont {R.}~\bibnamefont {Alert}},
  \bibinfo {author} {\bibfnamefont {N.~S.}\ \bibnamefont {Wingreen}}, \ and\
  \bibinfo {author} {\bibfnamefont {J.~W.}\ \bibnamefont {Shaevitz}},\
  }\href@noop {} {\bibfield  {journal} {\bibinfo  {journal} {Nature Physics}\
  }\textbf {\bibinfo {volume} {17}},\ \bibinfo {pages} {211} (\bibinfo {year}
  {2021})}\BibitemShut {NoStop}%
\bibitem [{\citenamefont {Doostmohammadi}\ \emph
  {et~al.}(2016{\natexlab{b}})\citenamefont {Doostmohammadi}, \citenamefont
  {Adamer}, \citenamefont {Thampi},\ and\ \citenamefont
  {Yeomans}}]{doostmohammadi2016stabilization}%
  \BibitemOpen
  \bibfield  {author} {\bibinfo {author} {\bibfnamefont {A.}~\bibnamefont
  {Doostmohammadi}}, \bibinfo {author} {\bibfnamefont {M.~F.}\ \bibnamefont
  {Adamer}}, \bibinfo {author} {\bibfnamefont {S.~P.}\ \bibnamefont {Thampi}},
  \ and\ \bibinfo {author} {\bibfnamefont {J.~M.}\ \bibnamefont {Yeomans}},\
  }\href@noop {} {\bibfield  {journal} {\bibinfo  {journal} {Nature
  communications}\ }\textbf {\bibinfo {volume} {7}},\ \bibinfo {pages} {1}
  (\bibinfo {year} {2016}{\natexlab{b}})}\BibitemShut {NoStop}%
\bibitem [{\citenamefont {Chat{\'e}}(2020)}]{chate2020dry}%
  \BibitemOpen
  \bibfield  {author} {\bibinfo {author} {\bibfnamefont {H.}~\bibnamefont
  {Chat{\'e}}},\ }\href@noop {} {\bibfield  {journal} {\bibinfo  {journal}
  {Annual Review of Condensed Matter Physics}\ }\textbf {\bibinfo {volume}
  {11}},\ \bibinfo {pages} {189} (\bibinfo {year} {2020})}\BibitemShut
  {NoStop}%
\bibitem [{\citenamefont {Simha}\ and\ \citenamefont
  {Ramaswamy}(2002)}]{simha2002hydrodynamic}%
  \BibitemOpen
  \bibfield  {author} {\bibinfo {author} {\bibfnamefont {R.~A.}\ \bibnamefont
  {Simha}}\ and\ \bibinfo {author} {\bibfnamefont {S.}~\bibnamefont
  {Ramaswamy}},\ }\href@noop {} {\bibfield  {journal} {\bibinfo  {journal}
  {Physical review letters}\ }\textbf {\bibinfo {volume} {89}},\ \bibinfo
  {pages} {058101} (\bibinfo {year} {2002})}\BibitemShut {NoStop}%
\bibitem [{\citenamefont {Srivastava}\ \emph {et~al.}(2016)\citenamefont
  {Srivastava}, \citenamefont {Mishra},\ and\ \citenamefont
  {Marchetti}}]{srivastava2016negative}%
  \BibitemOpen
  \bibfield  {author} {\bibinfo {author} {\bibfnamefont {P.}~\bibnamefont
  {Srivastava}}, \bibinfo {author} {\bibfnamefont {P.}~\bibnamefont {Mishra}},
  \ and\ \bibinfo {author} {\bibfnamefont {M.~C.}\ \bibnamefont {Marchetti}},\
  }\href@noop {} {\bibfield  {journal} {\bibinfo  {journal} {Soft matter}\
  }\textbf {\bibinfo {volume} {12}},\ \bibinfo {pages} {8214} (\bibinfo {year}
  {2016})}\BibitemShut {NoStop}%
\bibitem [{\citenamefont {Putzig}\ \emph {et~al.}(2016)\citenamefont {Putzig},
  \citenamefont {Redner}, \citenamefont {Baskaran},\ and\ \citenamefont
  {Baskaran}}]{putzig2016instabilities}%
  \BibitemOpen
  \bibfield  {author} {\bibinfo {author} {\bibfnamefont {E.}~\bibnamefont
  {Putzig}}, \bibinfo {author} {\bibfnamefont {G.~S.}\ \bibnamefont {Redner}},
  \bibinfo {author} {\bibfnamefont {A.}~\bibnamefont {Baskaran}}, \ and\
  \bibinfo {author} {\bibfnamefont {A.}~\bibnamefont {Baskaran}},\ }\href@noop
  {} {\bibfield  {journal} {\bibinfo  {journal} {Soft matter}\ }\textbf
  {\bibinfo {volume} {12}},\ \bibinfo {pages} {3854} (\bibinfo {year}
  {2016})}\BibitemShut {NoStop}%
\bibitem [{\citenamefont {Patelli}\ \emph {et~al.}(2019)\citenamefont
  {Patelli}, \citenamefont {Djafer-Cherif}, \citenamefont {Aranson},
  \citenamefont {Bertin},\ and\ \citenamefont
  {Chat{\'e}}}]{patelli2019understanding}%
  \BibitemOpen
  \bibfield  {author} {\bibinfo {author} {\bibfnamefont {A.}~\bibnamefont
  {Patelli}}, \bibinfo {author} {\bibfnamefont {I.}~\bibnamefont
  {Djafer-Cherif}}, \bibinfo {author} {\bibfnamefont {I.~S.}\ \bibnamefont
  {Aranson}}, \bibinfo {author} {\bibfnamefont {E.}~\bibnamefont {Bertin}}, \
  and\ \bibinfo {author} {\bibfnamefont {H.}~\bibnamefont {Chat{\'e}}},\
  }\href@noop {} {\bibfield  {journal} {\bibinfo  {journal} {Physical review
  letters}\ }\textbf {\bibinfo {volume} {123}},\ \bibinfo {pages} {258001}
  (\bibinfo {year} {2019})}\BibitemShut {NoStop}%
\bibitem [{\citenamefont {Shankar}\ and\ \citenamefont
  {Marchetti}(2019)}]{shankar2019hydrodynamics}%
  \BibitemOpen
  \bibfield  {author} {\bibinfo {author} {\bibfnamefont {S.}~\bibnamefont
  {Shankar}}\ and\ \bibinfo {author} {\bibfnamefont {M.~C.}\ \bibnamefont
  {Marchetti}},\ }\href@noop {} {\bibfield  {journal} {\bibinfo  {journal}
  {Physical Review X}\ }\textbf {\bibinfo {volume} {9}},\ \bibinfo {pages}
  {041047} (\bibinfo {year} {2019})}\BibitemShut {NoStop}%
\bibitem [{\citenamefont {Beris}\ \emph {et~al.}(1994)\citenamefont {Beris},
  \citenamefont {Edwards} \emph {et~al.}}]{beris1994thermodynamics}%
  \BibitemOpen
  \bibfield  {author} {\bibinfo {author} {\bibfnamefont {A.~N.}\ \bibnamefont
  {Beris}}, \bibinfo {author} {\bibfnamefont {B.~J.}\ \bibnamefont {Edwards}},
  \emph {et~al.},\ }\href@noop {} {\emph {\bibinfo {title} {Thermodynamics of
  flowing systems: with internal microstructure}}},\ \bibinfo {number} {36}\
  (\bibinfo  {publisher} {Oxford University Press on Demand},\ \bibinfo {year}
  {1994})\BibitemShut {NoStop}%
\bibitem [{\citenamefont {Marenduzzo}\ \emph {et~al.}(2007)\citenamefont
  {Marenduzzo}, \citenamefont {Orlandini},\ and\ \citenamefont
  {Yeomans}}]{marenduzzo2007hydrodynamics}%
  \BibitemOpen
  \bibfield  {author} {\bibinfo {author} {\bibfnamefont {D.}~\bibnamefont
  {Marenduzzo}}, \bibinfo {author} {\bibfnamefont {E.}~\bibnamefont
  {Orlandini}}, \ and\ \bibinfo {author} {\bibfnamefont {J.~M.}\ \bibnamefont
  {Yeomans}},\ }\href@noop {} {\bibfield  {journal} {\bibinfo  {journal}
  {Physical review letters}\ }\textbf {\bibinfo {volume} {98}},\ \bibinfo
  {pages} {118102} (\bibinfo {year} {2007})}\BibitemShut {NoStop}%
\bibitem [{\citenamefont {Giomi}\ \emph {et~al.}(2010)\citenamefont {Giomi},
  \citenamefont {Liverpool},\ and\ \citenamefont
  {Marchetti}}]{giomi2010sheared}%
  \BibitemOpen
  \bibfield  {author} {\bibinfo {author} {\bibfnamefont {L.}~\bibnamefont
  {Giomi}}, \bibinfo {author} {\bibfnamefont {T.~B.}\ \bibnamefont
  {Liverpool}}, \ and\ \bibinfo {author} {\bibfnamefont {M.~C.}\ \bibnamefont
  {Marchetti}},\ }\href@noop {} {\bibfield  {journal} {\bibinfo  {journal}
  {Physical Review E}\ }\textbf {\bibinfo {volume} {81}},\ \bibinfo {pages}
  {051908} (\bibinfo {year} {2010})}\BibitemShut {NoStop}%
\bibitem [{\citenamefont {Oza}\ and\ \citenamefont
  {Dunkel}(2016)}]{oza2016antipolar}%
  \BibitemOpen
  \bibfield  {author} {\bibinfo {author} {\bibfnamefont {A.~U.}\ \bibnamefont
  {Oza}}\ and\ \bibinfo {author} {\bibfnamefont {J.}~\bibnamefont {Dunkel}},\
  }\href@noop {} {\bibfield  {journal} {\bibinfo  {journal} {New Journal of
  Physics}\ }\textbf {\bibinfo {volume} {18}},\ \bibinfo {pages} {093006}
  (\bibinfo {year} {2016})}\BibitemShut {NoStop}%
\bibitem [{\citenamefont {Shankar}\ \emph {et~al.}(2018)\citenamefont
  {Shankar}, \citenamefont {Ramaswamy}, \citenamefont {Marchetti},\ and\
  \citenamefont {Bowick}}]{shankar2018defect}%
  \BibitemOpen
  \bibfield  {author} {\bibinfo {author} {\bibfnamefont {S.}~\bibnamefont
  {Shankar}}, \bibinfo {author} {\bibfnamefont {S.}~\bibnamefont {Ramaswamy}},
  \bibinfo {author} {\bibfnamefont {M.~C.}\ \bibnamefont {Marchetti}}, \ and\
  \bibinfo {author} {\bibfnamefont {M.~J.}\ \bibnamefont {Bowick}},\
  }\href@noop {} {\bibfield  {journal} {\bibinfo  {journal} {Physical review
  letters}\ }\textbf {\bibinfo {volume} {121}},\ \bibinfo {pages} {108002}
  (\bibinfo {year} {2018})}\BibitemShut {NoStop}%
\bibitem [{\citenamefont {DeCamp}\ \emph {et~al.}(2015)\citenamefont {DeCamp},
  \citenamefont {Redner}, \citenamefont {Baskaran}, \citenamefont {Hagan},\
  and\ \citenamefont {Dogic}}]{decamp2015orientational}%
  \BibitemOpen
  \bibfield  {author} {\bibinfo {author} {\bibfnamefont {S.~J.}\ \bibnamefont
  {DeCamp}}, \bibinfo {author} {\bibfnamefont {G.~S.}\ \bibnamefont {Redner}},
  \bibinfo {author} {\bibfnamefont {A.}~\bibnamefont {Baskaran}}, \bibinfo
  {author} {\bibfnamefont {M.~F.}\ \bibnamefont {Hagan}}, \ and\ \bibinfo
  {author} {\bibfnamefont {Z.}~\bibnamefont {Dogic}},\ }\href@noop {}
  {\bibfield  {journal} {\bibinfo  {journal} {Nature materials}\ }\textbf
  {\bibinfo {volume} {14}},\ \bibinfo {pages} {1110} (\bibinfo {year}
  {2015})}\BibitemShut {NoStop}%
\bibitem [{\citenamefont {Chung}(2010)}]{chung2010}%
  \BibitemOpen
  \bibfield  {author} {\bibinfo {author} {\bibfnamefont {T.}~\bibnamefont
  {Chung}},\ }\href@noop {} {\emph {\bibinfo {title} {Computational Fluid
  Dynamics}}}\ (\bibinfo  {publisher} {Cambridge University Press},\ \bibinfo
  {year} {2010})\BibitemShut {NoStop}%
\bibitem [{\citenamefont {Chardac}\ \emph {et~al.}(2021)\citenamefont
  {Chardac}, \citenamefont {Hoffmann}, \citenamefont {Poupart}, \citenamefont
  {Giomi},\ and\ \citenamefont {Bartolo}}]{chardac2021topology}%
  \BibitemOpen
  \bibfield  {author} {\bibinfo {author} {\bibfnamefont {A.}~\bibnamefont
  {Chardac}}, \bibinfo {author} {\bibfnamefont {L.~A.}\ \bibnamefont
  {Hoffmann}}, \bibinfo {author} {\bibfnamefont {Y.}~\bibnamefont {Poupart}},
  \bibinfo {author} {\bibfnamefont {L.}~\bibnamefont {Giomi}}, \ and\ \bibinfo
  {author} {\bibfnamefont {D.}~\bibnamefont {Bartolo}},\ }\href@noop {}
  {\bibfield  {journal} {\bibinfo  {journal} {Physical Review X}\ }\textbf
  {\bibinfo {volume} {11}},\ \bibinfo {pages} {031069} (\bibinfo {year}
  {2021})}\BibitemShut {NoStop}%
\bibitem [{\citenamefont {Jiang}\ and\ \citenamefont
  {Wang}(2019)}]{jiang2019trait}%
  \BibitemOpen
  \bibfield  {author} {\bibinfo {author} {\bibfnamefont {H.}~\bibnamefont
  {Jiang}}\ and\ \bibinfo {author} {\bibfnamefont {S.}~\bibnamefont {Wang}},\
  }\href@noop {} {\bibfield  {journal} {\bibinfo  {journal} {Physical Review
  Research}\ }\textbf {\bibinfo {volume} {1}},\ \bibinfo {pages} {033164}
  (\bibinfo {year} {2019})}\BibitemShut {NoStop}%
\bibitem [{\citenamefont {Vromans}\ and\ \citenamefont
  {Giomi}(2016)}]{vromans2016orientational}%
  \BibitemOpen
  \bibfield  {author} {\bibinfo {author} {\bibfnamefont {A.~J.}\ \bibnamefont
  {Vromans}}\ and\ \bibinfo {author} {\bibfnamefont {L.}~\bibnamefont
  {Giomi}},\ }\href@noop {} {\bibfield  {journal} {\bibinfo  {journal} {Soft
  matter}\ }\textbf {\bibinfo {volume} {12}},\ \bibinfo {pages} {6490}
  (\bibinfo {year} {2016})}\BibitemShut {NoStop}%
\bibitem [{\citenamefont {Tang}\ and\ \citenamefont
  {Selinger}(2017)}]{tang2017orientation}%
  \BibitemOpen
  \bibfield  {author} {\bibinfo {author} {\bibfnamefont {X.}~\bibnamefont
  {Tang}}\ and\ \bibinfo {author} {\bibfnamefont {J.~V.}\ \bibnamefont
  {Selinger}},\ }\href@noop {} {\bibfield  {journal} {\bibinfo  {journal} {Soft
  matter}\ }\textbf {\bibinfo {volume} {13}},\ \bibinfo {pages} {5481}
  (\bibinfo {year} {2017})}\BibitemShut {NoStop}%
\bibitem [{\citenamefont {Pearce}\ \emph {et~al.}(2021)\citenamefont {Pearce},
  \citenamefont {Nambisan}, \citenamefont {Ellis}, \citenamefont
  {Fernandez-Nieves},\ and\ \citenamefont {Giomi}}]{pearce2021orientational}%
  \BibitemOpen
  \bibfield  {author} {\bibinfo {author} {\bibfnamefont {D.~J.~G.}\
  \bibnamefont {Pearce}}, \bibinfo {author} {\bibfnamefont {J.}~\bibnamefont
  {Nambisan}}, \bibinfo {author} {\bibfnamefont {P.~W.}\ \bibnamefont {Ellis}},
  \bibinfo {author} {\bibfnamefont {A.}~\bibnamefont {Fernandez-Nieves}}, \
  and\ \bibinfo {author} {\bibfnamefont {L.}~\bibnamefont {Giomi}},\
  }\href@noop {} {\bibfield  {journal} {\bibinfo  {journal} {Physical Review
  Letters}\ }\textbf {\bibinfo {volume} {127}},\ \bibinfo {pages} {197801}
  (\bibinfo {year} {2021})}\BibitemShut {NoStop}%
\bibitem [{\citenamefont {Nejad}\ \emph {et~al.}(2021)\citenamefont {Nejad},
  \citenamefont {Doostmohammadi},\ and\ \citenamefont
  {Yeomans}}]{nejad2021memory}%
  \BibitemOpen
  \bibfield  {author} {\bibinfo {author} {\bibfnamefont {M.~R.}\ \bibnamefont
  {Nejad}}, \bibinfo {author} {\bibfnamefont {A.}~\bibnamefont
  {Doostmohammadi}}, \ and\ \bibinfo {author} {\bibfnamefont {J.~M.}\
  \bibnamefont {Yeomans}},\ }\href@noop {} {\bibfield  {journal} {\bibinfo
  {journal} {Soft Matter}\ }\textbf {\bibinfo {volume} {17}},\ \bibinfo {pages}
  {2500} (\bibinfo {year} {2021})}\BibitemShut {NoStop}%
\end{thebibliography}%

\newpage
\appendix
\section{Linear stability analysis with elastic stress}\label{app:A}
We present in this section the linear stability analysis of the uniform nematic state with $u_i=0$ and $\mathbf{Q_0}=\frac{S_0}{2}\begin{pmatrix} 1&0\\0&-1 \end{pmatrix}$, where $S_0=\sqrt{\frac{r-1}{r}}$. We omit the phenomenological surface tension term in this analysis and include a passive elastic stress 
\begin{equation}\label{eq:elastic_stress}
\sigma_{ij}^p = -\lambda H_{ij} 
\end{equation}
in the equation of the flow field. 
This analysis demonstrates that, in the linearized theory, the phenomenological surface tension term in the main text captures some of the effects of the passive elastic stress.
Appendix C provides a more general derivation of the surface tension term from elastic stress in a compressible fluid.

The linearized equation of motion of the order parameter is given by
\begin{equation}
\partial_t \delta Q_{ij}=\lambda \delta D_{ij}+
Q_{0,ik}\delta \omega_{kj}-\delta\omega_{ik}Q_{0,kj}+
\frac{1}{\gamma}\delta H_{ij},
\end{equation}
where
\begin{equation}
\delta H_{ij}=-4ArS_0\delta Q_{xx}Q_{0,ij}
+K\nabla^2\delta Q_{ij}
\end{equation}

\subsection{Incompressible fluid}
In an incompressible flow, the linearized equations of the flow are
\begin{equation}
\begin{aligned}
&\Gamma\delta u_i=-\partial_i\delta p + \partial_j (\delta \sigma_{ij}^a + \delta\sigma_{ij}^p)\\
&\partial_i\delta u_i=0,
\end{aligned}
\end{equation}
where $\delta\sigma_{ij}^a = \alpha\delta Q_{ij}$ is the active stress and $\delta\sigma_{ij}^p = -\lambda\delta H_{ij}$ is the passive elastic stress.
The flow field can be solved in Fourier space as
\begin{equation}
\delta u_i=\delta u_i^a + \delta u_i^p = \frac{P_{ik}}{\Gamma}(iq_j\delta \sigma_{kj}^a + iq_j\delta \sigma_{kj}^p),
\end{equation}
where the transverse projection operator is $P_{ik}=\delta_{ik}-\hat{q}_i \hat{q}_k$.

The components of the active contribution to the strain rate tensor are
\begin{equation}
\begin{aligned}
\delta D_{xx}^a &= \frac{-\alpha q^2}{2\Gamma}[2\hat{q}_x^2(1-\hat{q}_x^2 +\hat{q}_y^2)\delta Q_{xx}+2\hat{q}_x\hat{q}_y(1-2\hat{q}_x^2)\delta Q_{xy}]\\
&= \frac{-\alpha q^2}{2\Gamma}[\text{sin}^2 (2\phi)\delta Q_{xx}-\text{sin}(2\phi)\text{cos}(2\phi)\delta Q_{xy}],
\end{aligned}
\end{equation}
and
\begin{equation}
\begin{aligned}
\delta D_{xy}^a &=\frac{-\alpha q^2}{2\Gamma}[(1-4\hat{q}_x^2\hat{q}_y^2)\delta Q_{xy}-2\hat{q}_x\hat{q}_y(\hat{q}_x^2-\hat{q}_y^2)\delta Q_{xx}]\\
&= \frac{-\alpha q^2}{2\Gamma}[\text{cos}^2(2\phi)\delta Q_{xy}-\text{sin}(2\phi)\text{cos}(2\phi)\delta Q_{xx}].
\end{aligned}
\end{equation}
In the above equations, we have introduced $\hat{q}_x=\text{cos}(\phi)$, and $\hat{q}_y=\text{sin}(\phi)$, so that $\hat{q}_x^2-\hat{q}_y^2=\text{cos}(2\phi)$, $1-2\hat{q}_x^2=-\text{cos}(2\phi)$, $1-\hat{q}_x^2+\hat{q}_y^2=2\text{sin}^2(\phi)$.
The active contribution to the flow vorticity is
\begin{equation}
\begin{aligned}
\delta \omega_{xy}^a &= \frac{\alpha q^2}{2\Gamma}[(\hat{q}_y^2 -\hat{q}_x^2)\delta Q_{xy} + 2\hat{q}_x\hat{q}_y \delta Q_{xx}]\\
&= \frac{\alpha q^2}{2\Gamma}[-\text{cos}(2\phi)\delta Q_{xy}+\text{sin}(2\phi)\delta Q_{xx}].
\end{aligned}
\end{equation}

The components of the passive contribution to the strain rate tensor are
\begin{equation}
\begin{aligned}
\delta D_{xx}^p &= \frac{q^2}{\Gamma}[(-4\lambda ArS_0^2\hat{q}_x^2\hat{q}_y^2-2\lambda Kq^2\hat{q}_x^2\hat{q}_y^2)\delta Q_{xx}\\
&-\lambda Kq^2(\hat{q}_y^2-\hat{q}_x^2)\hat{q}_x\hat{q}_y \delta Q_{xy}]\\
&= \frac{q^2}{2\Gamma}[(-2\lambda ArS_0^2\text{sin}^2(2\phi)-\lambda Kq^2\text{sin}^2(2\phi))\delta Q_{xx}\\
&+ \lambda Kq^2\text{cos}(2\phi)\text{sin}(2\phi)\delta Q_{xy}],
\end{aligned}
\end{equation}
and
\begin{equation}
\begin{aligned}
\delta D_{xy}^p &= \frac{q^2}{2\Gamma}(\hat{q}_y^2-\hat{q}_x^2)[(-4\lambda ArS_0^2\hat{q}_x\hat{q}_y-2\lambda Kq^2 \hat{q}_x\hat{q}_y)\delta Q_{xx}\\
&+\lambda Kq^2(\hat{q}_x^2 -\hat{q}_y^2)\delta Q_{xy}]\\
&= \frac{q^2}{2\Gamma}\text{cos}(2\phi)[(2\lambda ArS_0^2\text{sin}(2\phi)+\lambda Kq^2\text{sin}(2\phi))\delta Q_{xx}\\
&- \lambda Kq^2 \text{cos}(2\phi)\delta Q_{xy}].
\end{aligned}
\end{equation}
The passive contribution to the vorticity is
\begin{equation}
\begin{aligned}
\delta \omega_{xy}^p &= \frac{q^2}{2\Gamma}[4\lambda ArS_0^2\hat{q}_x\hat{q}_y\delta Q_{xx}\\
&+\lambda Kq^2((\hat{q}_y^2 - \hat{q}_x^2)\delta Q_{xy}+2\hat{q}_x\hat{q}_y\delta Q_{xx})] \\
&= \frac{q^2}{2\Gamma}[2\lambda ArS_0^2\text{sin}(2\phi)\delta Q_{xx}\\
&+\lambda Kq^2(-\text{cos}(2\phi)\delta Q_{xy}+\text{sin}(2\phi)\delta Q_{xx})]
\end{aligned}
\end{equation}
We see that the passive parts of the strain rate tensor and vorticity contain terms proportional to $q^4$, which is an effective surface tension term.

Collecting the above results into Eq.(A2) then gives an explicit expression of the dynamical matrix for $\delta Q_{ij}$. One can check that the most unstable mode in an extensile active nematic system is the pure bend mode, corresponding to $\phi=0$ \cite{srivastava2016negative}.
In this case, the dynamics of $\delta Q_{xx}$ and $\delta Q_{xy}$ are decoupled as
\begin{equation}
\begin{aligned}
&\partial_t\delta Q_{xx}=(-\frac{2ArS_0^2}{\gamma}-\frac{K}{\gamma}q^2)\delta Q_{xx}\\
&\partial_t\delta Q_{xy}=[\frac{-\alpha\lambda q^2}{2\Gamma}-\frac{S_0\alpha q^2}{2\Gamma}-\frac{K}{\gamma}q^2 - \frac{Kq^4}{2\Gamma}\lambda(\lambda+S_0)]\delta Q_{xy}.
\end{aligned}
\end{equation}
$\delta Q_{xy}$ is unstable if $|\alpha|>\frac{2K\Gamma}{\gamma(\lambda+S_0)}$. The $O(q^4)$ term provides stability at short length scale. 

\subsection{Compressible fluid}
In a compressible fluid, the linearized equations of the flow are
\begin{equation}
\Gamma\delta u_i= \partial_j (\delta \sigma_{ij}^a + \delta\sigma_{ij}^p)    
\end{equation}
The active and passive contributions to the flow vorticity remain the same as in Eq.(A8) and Eq.(A11).
The active part of the strain rate tensor is given by
\begin{equation}
\delta D_{ij}^a = \frac{-\alpha q^2}{2\Gamma}\delta Q_{ij}.
\end{equation}
The components of the passive contribution to the strain rate tensor are given by
\begin{equation}
\delta D_{xx}^p = \frac{-1}{2\Gamma}(2\lambda ArS_0^2 q^2 \delta Q_{xx}+\lambda Kq^4 \delta Q_{xx}),
\end{equation}
and
\begin{equation}
\delta D_{xy}^p = \frac{-1}{2\Gamma}\lambda Kq^4 \delta Q_{xy}.
\end{equation}
Similarly to before, the stability of $\delta Q_{ij}$ can be analyzed by substituting these results into Eq.(A2).
The most unstable mode in a compressible fluid is also the pure bend mode, in which case, the dynamics of $\delta Q_{ij}$ is given by
\begin{equation}
\begin{aligned}
\partial_t \delta Q_{xx} &= [-\frac{2ArS_0^2}{\gamma}-\frac{\alpha\lambda q^2}{2\Gamma}-\frac{K}{\gamma}q^2\\
&-\frac{\lambda q^2}{2\Gamma}(2\lambda ArS_0^2 +\lambda Kq^2)] \delta Q_{xx}\\
\partial_t \delta Q_{xy} &= [\frac{-\alpha \lambda q^2}{2\Gamma}-\frac{S_0\alpha q^2}{2\Gamma}-\frac{K}{\gamma}q^2-\frac{Kq^4}{2\Gamma}\lambda(\lambda+S_0)] \delta Q_{xy}.
\end{aligned}
\end{equation}
The critical activity for instability is the same as the one in the incompressible case.

\section{Absence of arch solution in incompressible fluids}
\label{app:B}
In this appendix we recall the existence of arch solution in a compressible system and detail its absence for incompressible flows. We start with the equations in complex coordinates, as given in Eqs.~(\ref{eq:psi}) and (\ref{eq:u_cmplx}), with
the incompressibility constraint 
$\partial_{\bar{z}}\bar{u}+\partial_{z}u=0$.
The first two terms on the right hand side of Eq.~(\ref{eq:psi}) are the convective terms, which are followed by the terms for vorticity.

In the case of compressible flow the pressure gradient on the right hand side of Eq.~(\ref{eq:u_cmplx}) vanishes and one can immediately eliminate the flow velocity  from Eq.~(\ref{eq:psi}) using $u=2\alpha\partial_z\psi/\Gamma$. Substituting in Eq.~(\ref{eq:psi}) the solution for perfectly aligned arches with a constant phase gradient in the form given in Eq. (\ref{eq:arch_solution}), we then obtain Eq.~(\ref{eq:psi_arch}), where the advective term and those coupling to vorticity exactly cancel. The resulting equation can then be solved in a steady state, leading to the value of $S_k$ given in Eq.~(\ref{eq:arch}). 

The same arch structure is not a solution for the case of incompressible flow because in this case the nonlinear advective term and the term coupling to vorticity do not cancel. To see this, we  solve for the pressure gradient in Eq.(\ref{eq:u}), with the result 
\begin{equation}
\bm\nabla p=(\alpha \partial_x Q_{xx},0)^T + \bm\nabla h,
\end{equation}
where we assume the Q tensor is independent of $y$, and $h$ is a real harmonic function. The $T$ superscript denotes a vector transpose.
The flow is then given by
\begin{equation}
\begin{aligned}
\mathbf{u}=&\frac{\alpha}{\Gamma} (0,  \partial_x Q_{xy})^T -\bm\nabla h/\Gamma\\
=& \frac{2\alpha}{\Gamma}(\partial_z + \partial_{\bar{z}})(\psi -\bar{\psi})-\frac{1}{\Gamma}\partial_{\bar{z}}h,
\end{aligned}
\end{equation}
where the second line represents the flow in complex coordinates.
Using the ansatz in Eq.(\ref{eq:arch_solution}),
the advective terms yield
\begin{equation}
-u\partial_z \psi_k-\bar{u}\partial_{\bar{z}}\psi_k = \frac{ik}{\Gamma}(\partial_z h +\partial_{\bar{z}}h)\psi_k\;,
\end{equation}
and
\begin{equation}
\begin{aligned}
\partial_t\psi_k=& \frac{ik}{\Gamma}(\partial_z h +\partial_{\bar{z}}h)\psi_k-\frac{2\alpha}{\Gamma}k^2(\psi_k-\bar{\psi})\psi_k-\frac{2\alpha\lambda}{\Gamma}k^2\psi_k\\
&-\frac{A}{\gamma}(1-r+rS_k^2)\psi_k-\frac{4K}{\gamma}k^2\psi_k-\frac{16\kappa}{\gamma}k^4\psi_k\;.
\end{aligned}
\label{eq:psi_arch_incomp}
\end{equation}
In other words the contribution from the coupling to vorticity is no longer balanced by advection. The above equation cannot be satisfied by a harmonic function $h$. A solution would require, for instance, 
$\partial_z h =\frac{-2\alpha}{\Gamma}ik\psi_k$ which would give $\partial_{\bar{z}}\partial_z h\not=0.$ Therefore, the arch ansatz with constant phase gradient is not a steady state solution when the flow is  incompressible.

We note that in Ref.~\cite{nejad2021memory} by  focusing on the dynamics of the phase field $\theta(\bm r)$ and assuming constant order parameter magnitude, the authors showed that the vorticity terms can be balanced by the elastic stiffness term at low activity, $|\alpha|<\frac{2K\Gamma}{S_0\gamma}$, allowing arches solutions with spatially varying phase gradient.

\section{Effective surface tension}\label{app:C}

Here we show that the effective surface tension $\kappa$ arises from the elastic stress defined by Eq.~(\ref{eq:elastic_stress}) due to nematic distortions, namely 
\[
\sigma_{ij}^{p} \approx -\lambda K \nabla^2 Q_{ij}. 
\]
The overdamped flow velocity induced by this is therefore 
\[
u_i^{el} = -\frac{\lambda K}{\Gamma} \nabla^2\partial_k Q_{ik} 
\]
and contributes to the flow alignment of $Q$ tensor from Eq.~(\ref{eq:Qeq}) through an additional strain given by 
\[
D_{ij}^{el} = -\frac{\lambda K}{\Gamma} \nabla^2\left(\partial_{ik} Q_{jk} +\partial_{jk} Q_{ik}-\partial_{kl} Q_{kl}\delta_{ij}\right)
\]
where 
\[\partial_{kl} Q_{kl} = (\partial_x^2-\partial_y^2)Q_{xx}+2\partial_{xy}Q_{xy}.\]
From the symmetry of the $Q$ tensor, it quickly follows that this elastic strain is proportional to $\nabla^4 Q$. Consequently, the flow alignment due to this strain reduces to an effective surface tension. To see this, we evaluate the $xx$-component of $D_{ij}^{el}$ which is given by 
\begin{eqnarray*}
 D_{xx}^{el} &=& -\frac{\lambda K}{\Gamma} \nabla^2\left(2\partial_x^2Q_{xx}+2\partial_{xy}Q_{xy}-\partial_{kl} Q_{kl}\right)\\
 &=& -\frac{\lambda K}{\Gamma} \nabla^4 Q_{xx}
\end{eqnarray*}
Similarly, it follows that 
\[D_{yy}^{el} = -\frac{\lambda K}{\Gamma} \nabla^4 Q_{yy} = -D_{xx}^{el}\]
where $Q_{yy}=-Q_{xx}$. The $xy$ term is 
\begin{eqnarray*}
 D_{xy}^{el} &=& -\frac{\lambda K}{\Gamma} \nabla^2\left(\nabla^2 Q_{xy} +\partial_{xy}Q_{xx}+\partial_{xy}Q_{yy}\right)\\
 &=& -\frac{\lambda K}{\Gamma} \nabla^4 Q_{xy}
\end{eqnarray*}
Thus, to this leading order, the surface tension reduces to $\kappa = \lambda^2 K/\Gamma$. In general, however, it can also be taken as an independent parameter, and this is the route considered here. 

\end{document}